\documentclass[prl,twocolumn]{revtex4-1}
\usepackage[utf8]{inputenc}
\usepackage{hyperref}
\usepackage{graphicx}  % needed for figures
\usepackage{dcolumn}   % needed for some tables
\usepackage{bm}        % for math
\usepackage{amssymb,amsmath}   % for math
\usepackage{uri}

\usepackage[x11names, rgb, html]{xcolor}
\definecolor{sbase03}{HTML}{002B36}
\definecolor{sbase02}{HTML}{073642}
\definecolor{sbase01}{HTML}{586E75}
\definecolor{sbase00}{HTML}{657B83}
\definecolor{sbase0}{HTML}{839496}
\definecolor{sbase1}{HTML}{93A1A1}
\definecolor{sbase2}{HTML}{EEE8D5}
\definecolor{sbase3}{HTML}{FDF6E3}
\definecolor{syellow}{HTML}{B58900}
\definecolor{sorange}{HTML}{CB4B16}
\definecolor{sred}{HTML}{DC322F}
\definecolor{smagenta}{HTML}{D33682}
\definecolor{sviolet}{HTML}{6C71C4}
\definecolor{sblue}{HTML}{268BD2}
\definecolor{scyan}{HTML}{2AA198}
\definecolor{sgreen}{HTML}{859900}
\hypersetup{
  colorlinks = true,
  citecolor = {sred},
  urlcolor = {sblue}
}

\begin{document}
\title{Fundamental Bounds on First Passage Time Fluctuations for Currents}
\author{Todd R. Gingrich}
\email{toddging@mit.edu} 
\author{Jordan M.~Horowitz}
\affiliation{Physics of Living Systems Group, Department of Physics, Massachusetts Institute of Technology, 400 Technology Square, Cambridge, MA 02139}

\begin{abstract}
Current is a characteristic feature of nonequilibrium systems.
In stochastic systems, these currents exhibit fluctuations constrained by the rate of dissipation in accordance with the recently discovered thermodynamic uncertainty relation.
Here, we derive a conjugate uncertainty relationship for the first passage time to accumulate a fixed net current.
More generally, we use the tools of large-deviation theory to simply connect current fluctuations and first-passage-time fluctuations in the limit of long times and large currents.
With this connection, previously discovered symmetries and bounds on the large-deviation function for currents are readily transferred to first passage times.
\end{abstract}
\pacs{05.70.Ln,05.40.-a} 
% 05.70.Ln -> Non-Eq Thermodynamics
% 05.40.-a -> Fluctuation phenomena, random processes, noise, and Brownian motion
% 02.50.-r -> Probability theory, stochastic processes, and statistics

\maketitle

{\em Introduction.}---Thermodynamics constrains the fluctuations of nonequilibrium systems, as evidenced by a growing collection of universal predictions connecting dissipation to fluctuations. 
Examples include the fluctuation theorems~\cite{Evans1994,Gallavotti1995,Jarzynski1997,Kurchan1998,Crooks1999,Lebowitz1999,Seifert2005}, nonequilibrium fluctuation-dissipation theorems~\cite{Speck2006,Prost2009,Baiesi2009,Seifert2010,Chetrite2011,Baiesi2013,Maes2014}, and, more recently, the thermodynamic uncertainty relation~\cite{Barato2015,Gingrich2016,Maes2017}.
Remarkably, all these results can be viewed through one unifying lens, namely large-deviation theory~\cite{Touchette2009}.
In fact, over the past two decades this formalism has proven to be an essential tool for characterizing the dynamical fluctuations of nonequilibrium systems~\cite{Maes2007,Maes2008,Maes2008On,Chandler2010,Chetrite2013,Touchette2013,Bertini2015Large,Bertini2015Flows}.

Recently, these techniques have revealed a universal inequality between the far-from-equilibrium fluctuations in current\textemdash such as the flow of particles, energy or entropy\textemdash with the near-equilibrium fluctuations predicted by linear-response theory~\cite{Gingrich2016}. 
A useful corollary is the thermodynamic uncertainty relation~\cite{Barato2015}, which offers a fundamental trade-off between typical current fluctuations and dissipation~\footnote{In the long-time limit, the typical fluctuations exhibit small deviations about the steady-state current.}.
Specifically, a nonequilibrium Markov process generating an average time-integrated current $\left<J\right>$ during a long observation time $T_{\rm obs}$ has a variance ${\rm Var}(J)$ constrained by the mean entropy-production rate $\sigma$ (with Boltzmann's constant $k_{\rm B}=1$):
\begin{equation}
\frac{{\rm Var}(J)}{\langle J\rangle^2} \geq \frac{2}{T_{\rm obs} \sigma}.
\label{eq:uncertainty}
\end{equation}
Thus, reducing fluctuations comes with an energetic cost.

\begin{figure}[tbh]
\begin{center}
\includegraphics[width=0.47\textwidth]{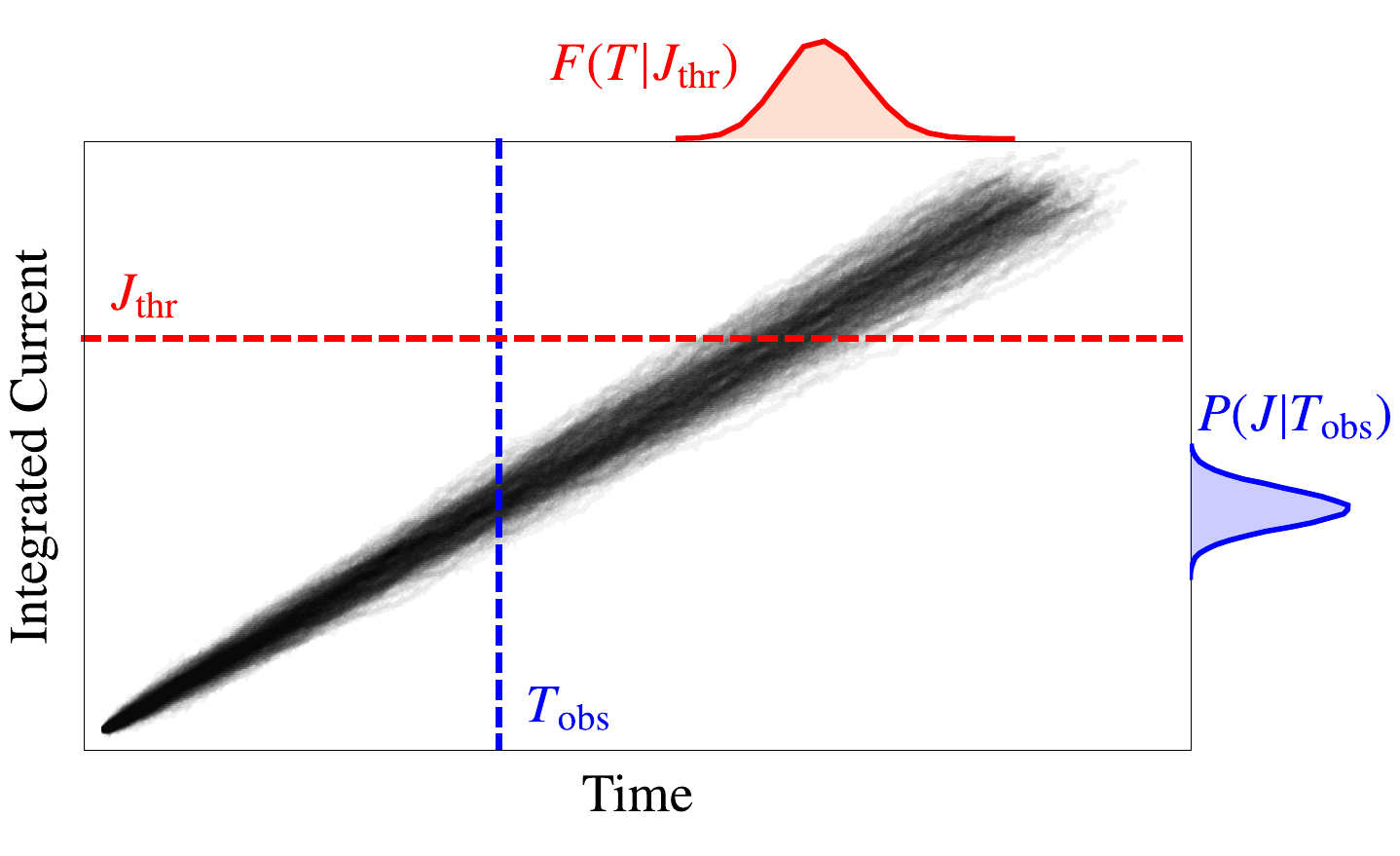}
\caption{
The distribution of integrated current $J$ for a long observation time $T_{\rm obs}$ and the distribution for first passage time $T$ to a large threshold current $J_{\rm thr}$ are two faces of the same distribution over trajectories. Hence, recent results describing the asymptotic form of the current distribution $P(J|T_{\rm obs})$ naturally yield corresponding results for the asymptotic form of the first passage time distribution $F(T|J_{\rm thr})$.
}
\label{fig:TrajFig}
\end{center}
\end{figure}

A significant body of recent work has analyzed such current fluctuations for a fixed observation time~\cite{Barato2015, Gingrich2016, Pietzonka2016Universal, Pietzonka2016MolecularMotors, Pietzonka2016Affinity, Polettini2016, Nyawo2016, Pietzonka2017, Pietzonka2017universal, Gingrich2017}.
In this Letter, we consider the complementary problem, analyzing the fluctuations of first passage times $T$ to reach a large threshold time-integrated  current $J_{\rm thr}$ (see Fig.~\ref{fig:TrajFig}).
We show that properties of the first passage time distribution for asymptotically large $J_{\rm thr}$ follow simply from knowledge of the current fluctuations.
This conjugate relationship between fixed-time and fixed-current trajectory ensembles mirrors the study of inverse or adjoint processes in queuing theory~\cite{Glynn1994, Russell1997, Duffy2004}, and it extends Garrahan's work on first passage time fluctuations of dynamical activity\textemdash a monotonically increasing counting variable~\cite{Budini2014, Garrahan2017}\textemdash to current variables which can grow or shrink.
By relating the conjugate problems, we are able to transform inequalities governing current fluctuations into associated inequalities for passage-time fluctuations, as well as offer fresh insight into recent predictions for entropy-production first passage times~\cite{Roldan2015, Saito2016, Neri2017, Pigolotti2017, Roldan2015}.
For instance, we show that the distribution for the time $T$ to first hit a large threshold current $J_{\rm thr}$ must satisfy a corresponding uncertainty relation:
\begin{equation}
  \frac{\text{Var}(T)}{\langle T\rangle^2} \geq \frac{2}{\langle T \rangle \sigma}.
  \label{eq:timeuncertainty}
\end{equation}

The two faces of the thermodynamic uncertainty relationship can be viewed as two ways to infer a bound on the entropy-production rate\textemdash one utilizing the current fluctuations in a fixed-time ensemble and the other utilizing the time fluctuations in a fixed-current ensemble.
Though these two sets of fluctuations contain equivalent information, we emphasize that the physical measurements are quite distinct.

{\it Setup.---} To make the notions concrete, we focus our presentation on nonequilibrium systems that can be modeled as Markov jump processes.
Specifically, we have in mind a mesoscopic system with states $i=1,\dots, N$, whose time-varying probability density ${\boldsymbol p}=\{p_i\}_{i=1}^N$ evolves according to the master equation $\dot{\boldsymbol p} = \mathbb{W} \boldsymbol{p}$,
where ${\mathbb W}_{ij}$ is the probability rate to transition from $j\to i$, and $-{\mathbb W}_{ii}=\sum_{j\neq i}{\mathbb W}_{ji}$ is the exit rate from $i$.
We assume that $\mathbb{W}$ is irreducible -- so that a unique steady-state exists -- and that every transition is reversible, that is ${\mathbb W}_{ij}\neq 0$ only when ${\mathbb W}_{ji}\neq 0$.
Thermodynamics enters by requiring transitions to satisfy local detailed balance.
The ratio of rates for each transition can then be identified with a generalized thermodynamic force $\mathcal{F}_{ij}=\ln({\mathbb W}_{ij}/{\mathbb W}_{ji})$~\footnote{The thermodynamic force may alternatively be defined in terms of the steady state density ${\boldsymbol \pi}$ as $\mathcal{F}_{ij} = \ln \left(\mathbb{W}_{ij} \pi_j / \mathbb{W}_{ji} \pi_i\right)$. These two definitions differ by the change in Shannon entropy $\ln\left(\pi_j / \pi_i\right)$ which averages to zero over a long trajectory.}, which quantifies the flow of free energy into the surrounding environment~\cite{Seifert2012}.

Fluctuating currents represent the net buildup of transitions between the system's mesoscopic states.
Indeed, in any given stochastic realization of our system's evolution there will be some random number of net transitions, or current, between every pair of states $j\to i$, which we label as $J_{ij}$.
Our interest though is in generalized currents obtained as superpositions of mesoscopic transitions, $J \equiv \sum_{i>j} d_{ij} J_{ij}$, where the $d_{ij}$ indicate how much a particular transition contributes.
Such generalized currents often represent a measurable global flow through the system, such as the ATP consumption throughout a biochemical network, or the net flow of heat between multiple thermal reservoirs~\cite{Seifert2012}.
A particularly important example is the fluctuating environmental entropy production $\Sigma$ obtained by choosing $d_{ij} = \ln \mathbb{W}_{ij} / \mathbb{W}_{ji}$.
Its average rate $\sigma=\lim_{T_{\rm obs}\to\infty}\langle\Sigma\rangle/T_{\rm obs}$ measures the time irreversibility of the dynamics.

\begin{figure}
\begin{center}
\includegraphics[width=0.46\textwidth]{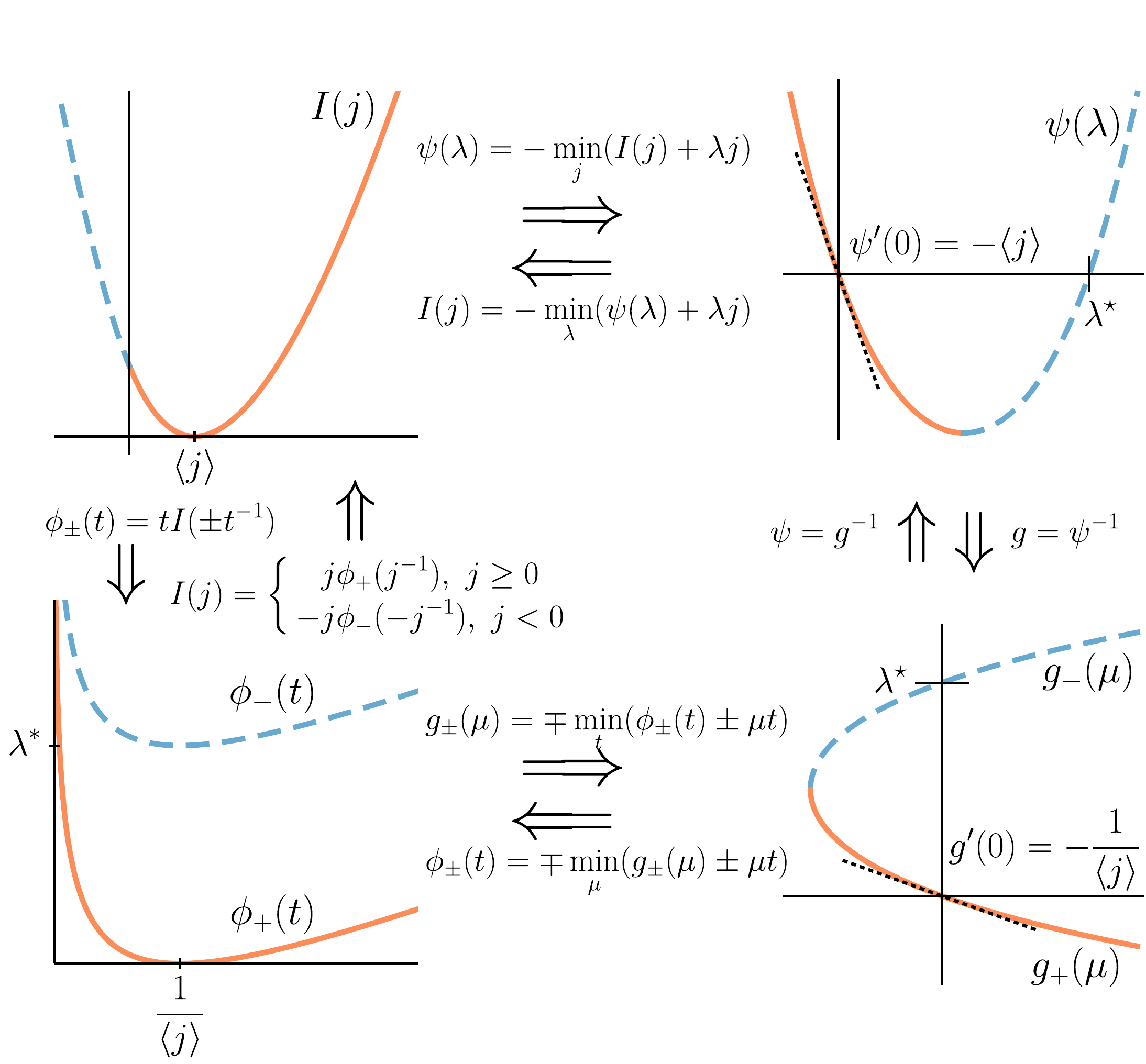}
\caption{
large-deviation rate functions (left) are related to SCGFs (right) by Legendre-Fenchel transform. 
Current statistics (top) and first passage time statistics (bottom) are connected by inversion.
Branches corresponding to positive currents are plotted with solid red lines, while the negative-current branches are plotted with dashed blue lines.
}
\label{fig:ConnectionsFig}
\end{center}
\end{figure}

For long observation times $T_{\rm obs}$, the probability of observing a current $J$ satisfies a large-deviation principle $P(J|T_{\rm obs})\asymp e^{-T_{\rm obs}I(J/T_{\rm obs})}$ with large-deviation rate function $I(j)$~\cite{Touchette2009}, where the lowercase letter $j \equiv J / T_{\rm obs}$ represents an intensive quantity.
The large-deviation function $I$ captures not just the typical fluctuations predicted by the central-limit theorem but also the relative likelihood of exponentially rare events.
A useful complementary characterization of the fluctuations is through the scaled cumulant generating function (SCGF) $\psi(\lambda)=\lim_{T_{\rm obs}\to\infty}(1/T_{\rm obs})\ln\langle e^{-\lambda J}\rangle$, with the expectation taken over trajectories of length $T_{\rm obs}$.
Derivatives of $\psi$ at the origin encode all the long-time current cumulants.
The pair $I$ and $\psi$ are intimately related through the Legendre-Fenchel transform, as graphically illustrated in Fig.~\ref{fig:ConnectionsFig}~\cite{Touchette2009}.

Universal symmetries and bounds on $I$  (commensurately $\psi$) have refined our understanding of the thermodynamics of nonequilibrium systems.
In the following, we develop a complementary point of view based on current first passage times. 

{\em First passage time fluctuations for large current.---}
We now consider a large (in magnitude) fixed amount of accumulated current $J_{\rm thr}$ and seek the time at which that threshold current is first reached.
As seen in Fig.~\ref{fig:TrajFig}, the mean first passage time scales extensively with the magnitude of $J_{\rm thr}$, suggesting a large-deviation form for the first passage time distribution $F(T|J_{\rm thr})$.
We note, however, that $J_{\rm thr}$ can be either positive or negative, and introduce two different rate functions, $\phi_+(t)$ and $\phi_-(t)$, to handle these cases:
\begin{equation}
F(T|J_{\rm thr})\asymp \begin{cases}
e^{-J_{\rm thr}\phi_+(T/J_{\rm thr})},& J_{\rm thr} > 0\\
e^{J_{\rm thr}\phi_-(-T/J_{\rm thr})},& J_{\rm thr} < 0.
\end{cases}
\end{equation}
Correspondingly, there are now two different SCGFs $g_{\pm}(\mu) = \lim_{J_{\rm thr} \to \pm \infty} (1 / J_{\rm thr}) \ln \left<e^{-\mu T}\right>$, with the expectation computed over trajectories having a fixed time-integrated current $J_{\rm thr}$.
Without loss of generality, we assume a choice of $\left\{d_{ij}\right\}$ such that $\left<J\right> > 0$.
In this case, the $+$ subscript corresponds to branches quantifying typical (positive-current) fluctuations and the $-$ subscript corresponds to rare (negative-current) branches.
It is useful to also split $\psi$ into two branches, $\psi_+$ with negative slope and $\psi_-$ with positive slope (see Fig.~\ref{fig:ConnectionsFig}).
Our central result is that the large deviations in scaled first passage times $t\equiv T/|J_{\rm thr}|$ are completely determined by the large-deviation functions for current fluctuations:
\begin{equation}
\phi_{\pm}(t)=tI(\pm 1/t),\qquad g_{\pm}(\mu)=\psi_{\pm}^{-1}(\mu).
\label{eq:phiI}
\end{equation}
Analogous relations have appeared for counting variables~\cite{Glynn1994, Russell1997, Duffy2004, Garrahan2017} and for entropy-production fluctuations~\cite{Saito2016}, but we show these connections are, in fact, more general and extend to all currents.
Thus, all known properties of $I$\textemdash most notably, symmetries and bounds\textemdash can naturally be translated to $\phi$.

Here, we offer a heuristic argument for Eq.~\eqref{eq:phiI} assuming positive current.
A sketch of a proof is included at the end of the Letter, and a more detailed proof is provided in the Supplemental Material (SM).
To start, we write ${\mathcal P}(\gamma)$ to denote the probability distribution for a mesoscopic trajectory $\gamma$\textemdash that is a sequence of states visited by the system and their jump times.
Then the likelihood of a large first passage time $T=tJ$ to a large current $J$ can be conveniently expressed as 
\begin{equation}
P(T=tJ)=\int d\gamma\, \delta(T-tJ) {\mathcal P}(\gamma),
\end{equation}
where the integral is over all trajectories.
However, the only trajectories that can contribute to this integral have current $J$.
Furthermore, large current can only be attained after a long time.
Taken together these observations suggest we can replace ${\mathcal P}$ with the large-deviation form for large $T$~\footnote{In passing from $\mathcal{P}(\gamma)$ to $e^{-T I(J/T)}$ we must recognize that $I(J/T)$ measures the asymptotic probability of a trajectory with net current $J$ in time $T$, including trajectories which have already hit $J$ at earlier times. Provided $\left|\left<J\right>\right| > 0$, the probability that the trajectory is making a first passage dwarfs the probability of repeated passages in the large $J$ limit.}:
\begin{equation}
P(T=tJ)\asymp\int dJ\, \delta(T-tJ) e^{-TI(J/T)}=e^{-JtI(1/t)},
\end{equation}
which implies $\phi_+(t) = t I(1/t)$, and $g_+(\mu)$ follows by Legendre-Fenchel transform.
Put simply, switching from current to first passage time is a change of variables where we replace current by its inverse.

We now turn to the implications of Eq.~\eqref{eq:phiI}.
For any generalized current, its  long-time fluctuations are constrained by the entropy-production rate via Eq.~\eqref{eq:uncertainty}.
This constraint actually follows from an inequality on the large-deviation rate function,
\begin{equation}
I(j)\le \frac{(j-\langle j\rangle)^2}{4\langle j\rangle^2}\sigma \equiv I_{\rm bnd}(j).
\label{eq:LDFbound}
\end{equation}
Translating to first passage time fluctuations, we have
\begin{equation}
\phi_+(t)\le \frac{(t-\langle t\rangle)^2}{4t}\sigma \equiv \phi_{\rm bnd}(t),
\label{eq:FPTbound}
\end{equation}
after noting that the typical behavior $\langle j\rangle=1/\langle t\rangle$ does not depend on the choice of ensemble -- fixed $T_{\rm obs}$ versus fixed $J_{\rm thr}$.
Equation~\eqref{eq:timeuncertainty} follows since the large $J_{\rm thr}$ variance is computed in terms of derivatives of the large-deviation function as $\text{Var}(T) = J_{\rm thr} / \phi_+''(\left<t\right>)$~\cite{Touchette2009}.
Thus, dissipation is a fundamental constraint to controlling first passage time fluctuations as well as current fluctuations.

Together Eqs.~\eqref{eq:LDFbound} and \eqref{eq:FPTbound} point to a remarkable property of the stochastic evolution of currents, which is best appreciated by normalizing the large-deviation forms $e^{-T_{\rm obs}I_{\rm bnd}(j)}$ and $e^{-J_{\rm thr}\phi_{\rm bnd}(t)}$.
For currents, we have a Gaussian distribution
\begin{equation}
P_{\rm bnd}(j) = \sqrt{\frac{T_{\rm obs}\sigma }{4 \pi \left<j\right>^2}} \exp\left[-\frac{T_{\rm obs} (j - \left<j\right>)^2 \sigma}{4 \left<j\right>^2}\right],
\label{eq:jbounding}
\end{equation}
whereas the first passage time distribution is an inverse Gaussian
\begin{equation}
F_{\rm bnd}(t) = \sqrt{\frac{J_{\rm thr}\sigma \left<t\right>^2 }{4 \pi t^3}} \exp\left[-\frac{J_{\rm thr}\left(t - \left<t\right>\right)^2 \sigma}{4 t}\right].
\label{eq:tbounding}
\end{equation}
Remarkably, these are the distributions we would have predicted if we had simply treated the evolution of the current as a one-dimensional diffusion process with constant drift $\langle j\rangle$ and diffusion coefficient $\sigma/\langle j\rangle^2$~\cite{Karlin}.
This observation suggests that while the precise dynamics of the currents is generally complex, there is a simple auxiliary diffusion process that constrains it, reminiscent of the universal form observed for the stochastic evolution of the entropy production as a drift-diffusion process~\cite{Seifert2005,Pigolotti2017}.

{\em First passage time fluctuations for negative current and the fluctuation theorem.---} We have focused primarily on first passage times to reach a (typical) positive current.
We can also consider the first passage time to the exponentially suppressed negative currents that arise due to trajectories that appear to run backwards in time.
The distribution for the time to reach $J_{\rm thr} < 0$ scales according to $\phi_-(t)$, which can be related to $\psi_-(\lambda)$ (see Fig.~\ref{fig:ConnectionsFig}).
This connection is especially interesting when $\psi$ posses a symmetry that relates its two branches $\psi_+$ and $\psi_-$, because this naturally translates to a relationship between $\phi_+$ and $\phi_-$.

Generically, $\psi_-$ vanishes at some $\lambda^*$. For certain currents it also satisfies $\psi_+(\lambda) = \psi_-(\lambda^* - \lambda)$.
As an example, the fluctuation theorem implies such a symmetry with $\lambda^*=1$ for the entropy production (itself a generalized current)~\cite{Lebowitz1999}.
Symmetry of $\psi$ yields a corresponding symmetry in $g_\pm$: $g_+(\mu) = -g_-(\mu) + \lambda^*$.
Taking the Legendre-Fenchel transform gives
\begin{equation}
\phi_+(t) = \phi_-(t) - \lambda^*,
\label{eq:phi+phi-}
\end{equation}
indicating that $\phi_+$ and $\phi_-$ differ by a constant offset when the SCGF symmetry is present.
Equation~\eqref{eq:phi+phi-} must be interpreted carefully, as it compares large-deviation functions for two different distributions.
Typically, large-deviation rate functions are shifted such that their minimum equals zero. In this case, a symmetrical $\psi$ implies that $\phi_+$ and $\phi_-$ are identical, and the large-current first passage time distribution $F(T|J_{\rm thr})$ is the same for both positive and negative $J_{\rm thr}$.
While the constant offset in Eq.~\eqref{eq:phi+phi-} does not affect the form of $F(T|J_{\rm thr})$, it reflects the fact that the probability of reaching $\left|J_{\rm thr}\right|$ exceeds that of reaching $-\left|J_{\rm thr}\right|$ by a factor of $e^{\lambda^* \left|J_{\rm thr}\right|}$.
Using the same methods as those in this Letter, Saito and Dhar reached similar conclusions for the case that the generalized current is the entropy production~\cite{Saito2016}, and Neri {\it et al.}\ have proven a corresponding fluctuation theorem for entropy production stopping times using Martingale theory~\cite{Neri2017}.
Our result, Eq.~\eqref{eq:phi+phi-}, extends more generally to any current satisfying a SCGF symmetry about $\lambda^*$, including the example of the next section.

\begin{figure}
\begin{center}
\includegraphics[width=0.47\textwidth]{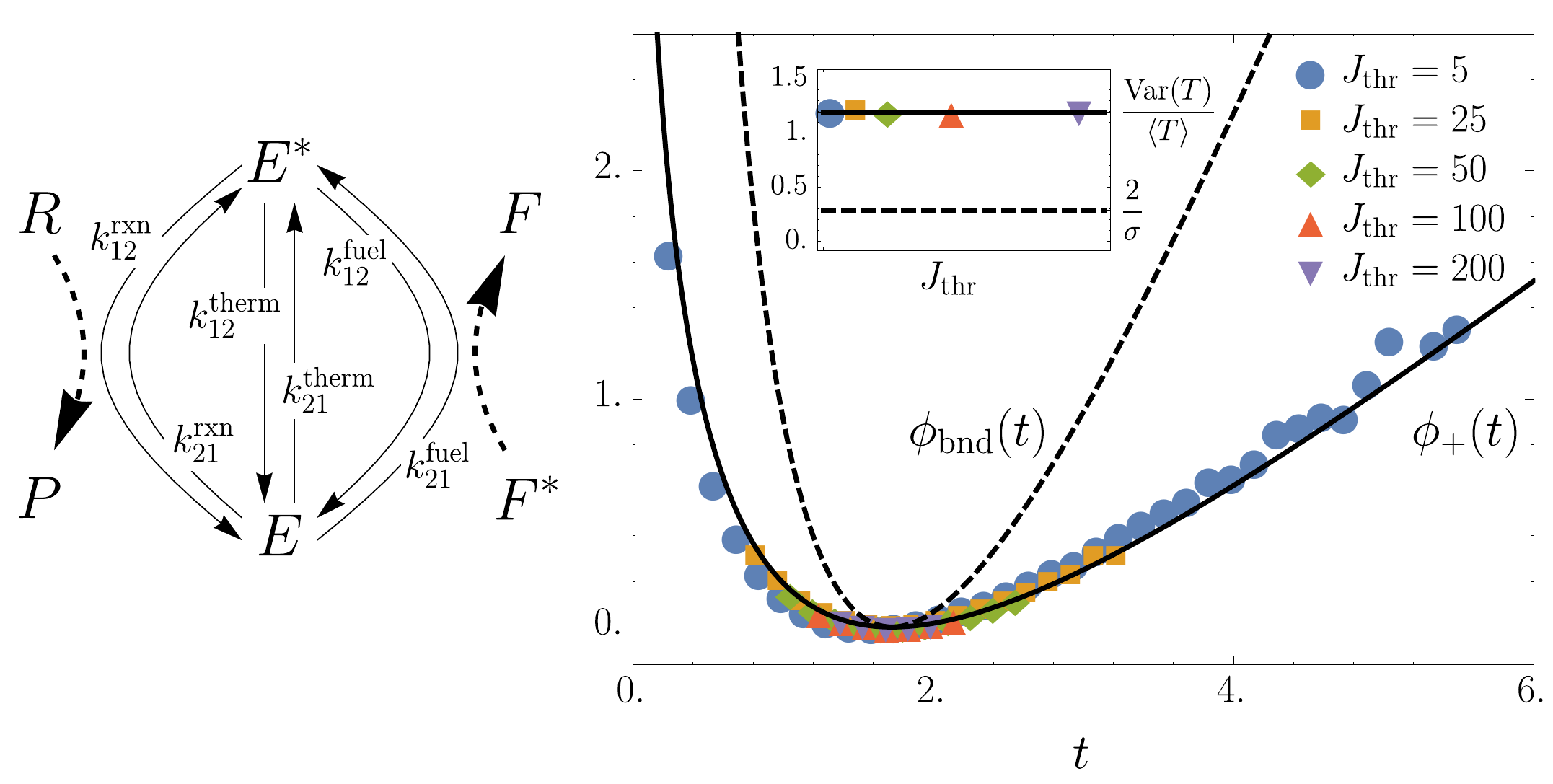}
\caption{
Markov model for the conversion of a reactant $R$ to product $P$ mediated by enzyme $E$.
The large-deviation function for the time to reach a particular net current from $R$ to $P$, $\phi_+(t)$, is bounded by $\phi_{\rm bnd}(t)$. Additionally, $\phi_+(t)$ is inferred from numerical sampling of $10^6$ trajectories for various choices of $J_{\rm thr}$ using rates: $k_{12}^{\rm rxn} = 2,\ k_{21}^{\rm rxn} = 0.1,\ k_{12}^{\rm therm} = 0.3,\ k_{21}^{\rm therm} = 0.001,\ k_{12}^{\rm fuel} = 0.001,\ k_{21}^{\rm fuel} = 1$.
}
\label{fig:LDFFig}
\end{center}
\end{figure}

{\em Illustrative example.---} To demonstrate the bounds in a more explicit context, we solve for the large-deviation behavior of a minimal model for an enzyme-mediated reaction from reactant $R$ to product $P$.
The enzyme can be either in a ground state $E$ or an activated state $E^*$, and the $E\leftrightarrow E^*$ transformations proceed via one of three pathways: (1) the enzyme exchanges heat with a thermal bath, (2) the enzyme accepts free energy by converting an activated fuel molecule $F^*$ into a deactivated form $F$, or (3) the activated enzyme converts $R \rightarrow P$.
Each of these pathways proceeds forward or backward, as depicted in Fig.~\ref{fig:LDFFig}, with six rate constants defining the model.
We follow the net transformations of $R$ into $P$ as the accumulated current $J$, so the first passage time can be interpreted as the time to generate $J$ product molecules.

The analytical solution of this model using standard methods is outlined in the SM.
Figure~\ref{fig:LDFFig} graphically shows the large-deviation function bound, Eq.~\eqref{eq:FPTbound}, as well as the uncertainty bound, Eq.~\eqref{eq:timeuncertainty} (see inset).
The analytical calculations are supplemented by trajectory sampling with finite $J_{\rm thr}$, the results of which are plotted with colored markers in Fig.~\ref{fig:LDFFig}.
Motivated by the $t^{-3/2}$ prefactor in Eq.~\eqref{eq:tbounding}, we extract estimates for $\phi_+(t)$ from the sampled trajectories by first approximating $F(T|J_{\rm thr})$ with a histogram and then computing
\begin{equation}
\phi_+^{\rm est}(t) = -\frac{1}{J_{\rm thr}} \left(\ln F(tJ_{\rm thr}|J_{\rm thr}) + \frac{3}{2} \ln t\right) + C_{\rm off},
\end{equation}
where $C_{\rm off}$ is a constant offset used to set the minimum of $\phi_+^{\rm est}$ to zero.
We observe that the large-deviation form (and, consequently, the thermodynamic uncertainty relation) remain valid even for small $J_{\rm thr}$.

{\em Conclusion.---} In the large-deviation limit, we have shown that current fluctuations with fixed observation time are intimately related to the fluctuations in first passage times to large current.
As a result, we have seen how the thermodynamic uncertainty relation and the fluctuation theorem for entropy production naturally lead to a universal symmetry and bounds on first passage time fluctuations.
Tighter-than-quadratic bounds on current large-deviation fluctuations~\cite{Pietzonka2016Universal, Pietzonka2016Affinity, Polettini2016} also readily translate to corresponding first passage time bounds.

Practically, we anticipate that it will be useful to convert between fixed-time and fixed-current ensembles since some experiments are more naturally suited to one than the other.
For example, imagine we seek a dissipation bound for the enzyme-mediated reaction in Fig.~\ref{fig:LDFFig}.
Fluctuations in product formation after time $T_{\rm obs}$ could be measured spectroscopically, assuming Beer's law and a calibrated mapping from fluorescence intensity to product concentration.
But the fixed $J_{\rm thr}$ ensemble offers an advantage.
By measuring first passage time fluctuations to reach a fixed fluorescence intensity, the mapping between fluorescence and concentration could be avoided altogether.
More ambitiously, we expect the fluctuating time ensemble to be a natural way to analyze the role of dissipation in Brownian clocks~\cite{Qian2000,Cao2015,Barato2016,Barato2017, Ray2017}.

{\em Sketch of a proof for Eq.~\eqref{eq:phiI}.---} 
The main result, Eq.~\eqref{eq:phiI}, consists of two relations: one connects the large-deviation rate function $I$ with $\phi_{\pm}$, the other connects $\psi$ with $g$.
Here we sketch a proof of $g_{\pm}(\mu) = \psi_{\pm}^{-1}(\mu)$.
The relationship between $I$ and $\phi_{\pm}$ follows by applying the G{\"a}rtner-Ellis theorem to compute $I$ from $\psi$ and $\phi_{\pm}$ from $g_{\pm}$.
More details are presented in the SM.

The basic strategy is to express both $g$ and $\psi$ in terms of spectral properties of a tilted rate matrix $\mathbb{W}(\lambda)$, whose elements are given by $\mathbb{W}_{ij}(\lambda) = \mathbb{W}_{ij} e^{-\lambda d_{ij}}$.
The first half of this connection is well known; the largest eigenvalue of $\mathbb{W}(\lambda)$ is the SCGF $\psi(\lambda)$.~\cite{Lebowitz1999}.
Expressing $g$ in terms of the tilted rate matrix requires a slightly more involved calculation following the general strategy of~\cite{Saito2016,Garrahan2017}.

Let $F_{ij}(T|J)$ be the distribution of times $T$ to first accumulate $J$ current with a jump to $i$, conditioned upon a start in $j$.
We connect $F_{ij}$ to the transition probability $P_{ij}(J, T)$ to go from $j \to i$ in time $T$, having accumulated current $J$ via the renewal equation: ${\bf P}(J,T)=\int_0^T dt\, {\bf P}(0,T-t)\cdot {\bf F}(t|J)$, written in matrix notation.
The convolution is simplified by Laplace transform (denoted with a tilde) to convert from $T$ to $\mu$, ultimately yielding $e^{-J g_\pm(\mu)} \asymp \langle e^{-\mu T}\rangle = \tilde{F}(\mu|J) \asymp {\bf \tilde{P}}(J, \mu)$.
Furthermore, ${\bf \tilde{P}}(J, \mu)$ can be expressed in terms of the tilted rate matrix via an inverse Laplace transform of ${\bf \widehat{\tilde{P}}}(\lambda, \mu) = 1 / (\mathbb{W}(\lambda) - \mu \mathbb{I})$, where the caret denotes a Laplace transform from $J$ to $\lambda$.
Using complex analysis to perform the inverse transform, we obtain $e^{-J g_\pm(\mu)} \asymp e^{\bar{\lambda}J}$, where $\bar{\lambda} = \psi_+^{-1}(\mu)$ for $J>0$ and $\bar{\lambda} = \psi_-^{-1}(\mu)$ for $J<0$.
Hence, $g_\pm$ and $\psi_\pm$ are inverses.

\begin{acknowledgments}
We gratefully acknowledge the Gordon and Betty Moore Foundation for supporting TRG and JMH as Physics of Living Systems Fellows through Grant GBMF4513.
\end{acknowledgments}

\bibliography{refs}

\newpage
\section{Supplemental Material}
\section{Derivations of main result}
The main result of the main text, Eq.\ (4), consists of two relations: one connects the large-deviation rate function $I$ with $\phi_{\pm}$, the other connects $\psi$ with $g$.
We first prove $g_{\pm}(\mu) = \psi_{\pm}^{-1}(\mu)$.
The relationship between $I$ and $\phi_{\pm}$ follows by applying the G{\"a}rtner-Ellis theorem to compute $I$ from $\psi$ and $\phi_{\pm}$ from $g_{\pm}$.

\subsection{Scaled cumulant generating functions $g$ and $\psi$ are inverses}
The basic strategy is to express both $g$ and $\psi$ in terms of spectral properties of a tilted rate matrix $\mathbb{W}(\lambda)$, whose elements are given by $\mathbb{W}_{ij}(\lambda) = \mathbb{W}_{ij} e^{-\lambda d_{ij}}$.
The first half of this connection is well-known; starting with initial density $\boldsymbol{\rho}$, the generating function for currents is obtained by the averaging over trajectories of length $T$ as $\left<e^{-\lambda J}\right> = \mathbf{1} \cdot e^{\mathbb{W}(\lambda) T} \cdot \boldsymbol{\rho}$, where $\mathbf{1} = \left\{1, \hdots, 1\right\}$~\cite{Lebowitz1999}.
It follows that the largest eigenvalue of $\mathbb{W}(\lambda)$ is the scaled cumulant generating function $\psi(\lambda)=\lim_{T\to\infty}(1/T)\ln \left<e^{-\lambda J}\right>$.

Expressing $g$ in terms of $\mathbb{W}(\lambda)$ requires a slightly more involved calculation.
We follow the general strategy of~\cite{Saito2016,Garrahan2017}.
First, we recall that $g$ is naturally expressed in terms of the Laplace transform of the first passage time distribution ${\tilde F}(\mu|J) \equiv \int_0^\infty dT e^{-\mu T} F(T|J)$ as
\begin{equation}\label{eq:g}
g_\pm(\mu)=\lim_{J\to\pm\infty}(1/J)\ln{\tilde F}(\mu|J).
\end{equation}
Thus, our goal is to express the large $J$ asymptotics of ${\tilde F}$ in terms of $\mathbb{W}(\lambda)$.

To this end, we introduce $F_{ij}(T|J)$ as the distribution of times to first reach $J$ current by a transition to $i$, given a start in $j$.
We connect $F_{ij}$ to the transition probability $P_{ij}(J, T)$ to go from $j \to i$ in time $T$, having accumulated current $J$ via the renewal equation:
\begin{equation}
P_{ij}(J,T)=\int_0^T dt\, \sum_{k} P_{ik}(0,T-t) F_{kj}(t|J).
\end{equation}
The convolution is made simpler by performing the Laplace transform (denoted with a tilde) to convert from $T$ to conjugate field $\mu$.
After minor rearrangement, the Laplace-transformed renewal equation leads to
\begin{align}
{\tilde F}(\mu|J) &= \mathbf{1}\cdot\tilde{\mathbf F}(\mu|J)\cdot\boldsymbol{\rho} \\
&= {\bf 1}\cdot {\bf \tilde{P}}(0,\mu)^{-1}\cdot{\bf \tilde{P}}(J,\mu)\cdot {\boldsymbol \rho},
\label{eq:conv}
\end{align}
where $\tilde{\mathbf{F}}$ and $\tilde{\mathbf{P}}$ are matrices with $ij$ matrix elements $\tilde{F}_{ij}$ and $\tilde{P}_{ij}$, respsectively.
The only term that contributes for large $J$ is ${\bf \tilde{P}}(J,\mu)$, which we analyze by taking an additional (two-sided) Laplace transform (denoted with a caret), this time a transform that converts from $J$ to a conjugate field $\lambda$:
\begin{align}
{\bf \widehat{\tilde{P}}}(\lambda,\mu)&=\int_{-\infty}^\infty dJ\, e^{-\lambda J}\int_{0}^\infty dT\, e^{-\mu T}{\bf P}(J,T).
\end{align}
By first performing the integral over $J$, we obtain
\begin{equation}
{\bf \widehat{\tilde{P}}}(\lambda,\mu)=\int_{0}^\infty dT\, e^{-(\mu {\mathbb I}-{\mathbb W}(\lambda))T}=\frac{1}{{\mathbb W}(\lambda)-\mu {\mathbb I}}.
\end{equation}
The integral is convergent only in the region $\psi_+^{-1}(\mu)<\lambda<\psi_-^{-1}(\mu)$.
We obtain ${\bm \tilde{P}}(J, \mu)$ by using a complex integral to invert the two-sided Laplace transform:
\begin{equation}
{\bf \tilde{P}}(J, \mu) = \frac{1}{2 \pi i} \int_C d\lambda\, {\bf \widehat{\tilde{P}}}(\lambda, \mu) e^{\lambda J}  = \frac{1}{2 \pi i}\int_C d\lambda\, \frac{e^{\lambda J}}{\mathbb{W}(\lambda) - \mu \mathbb{I}}.
\label{eq:contourintegral}
\end{equation}
The contour $C$ is chosen to be an infinite semicircle centered at a value of $\lambda$ chosen to fall inside the region of convergence.
So that the contour integral along the semicircular arc vanishes, $C$ must enclose the right half plane for $J<0$ or the left half plane for $J>0$.
The integral can then be performed using the residue theorem.
The asymptotic form for large $J$ is determined by the dominant pole, which comes from the the largest eigenvalue $\psi(\lambda)$ of ${\mathbb W}(\lambda)$.
 Hence, ${\bf \tilde{P}}(J, \mu) \asymp e^{\bar{\lambda} J}$, where $\bar{\lambda} = \psi_+^{-1}(\mu)$ for $J > 0$ and $\bar{\lambda} = \psi_-^{-1}(\mu)$ for $J < 0$.
Using Eq.~\eqref{eq:conv}, we get the large $J$ asymptotic scaling of the Laplace-transformed first-passage-time distribution, $\tilde{F}(\mu | J) \asymp e^{\bar{\lambda} J}$, and from Eq.~\eqref{eq:g} the SCGF $g_\pm(\mu) = \bar{\lambda} = \psi_\pm^{-1}(\mu)$.
We see that $\psi$ and $g$ are indeed inverses.

\subsection{Large-deviation rate functions are related by $\phi_{\pm}(t) = t I(\pm 1 / t)$}

By the G{\"a}rtner-Ellis theorem, $\phi_{\pm}$ and $g_\pm$ are related by a Legendre-Fenchel transform~\cite{Touchette2009}.
Hence,
\begin{align}
\nonumber \phi_{\pm}(t) &= \mp \min_\mu (g_\pm(\mu) \pm \mu t)\\
&= \mp g_{\pm}(\tilde{\mu}) - \tilde{\mu} t, \text{    with }g_{\pm}'(\tilde{\mu}) = \mp t,
\label{eq:phitransform}
\end{align}
where $\tilde{\mu}$ is the exponential bias that renders $t = T / |J_{\rm thr}|$ typical.
Similarly, in the fluctuating current ensemble, we define the exponential bias $\tilde{\lambda}$ that renders $j = J / T_{\rm obs}$ typical.
The Legendre-Fenchel transform relates $I$ to $\psi$ in terms of this $\tilde{\lambda}$:
\begin{align}
\nonumber I(j) &= -\min_\lambda (\psi(\lambda) + \lambda j)\\
&= - \psi(\tilde{\lambda}) - \tilde{\lambda} j, \text{     with }\psi'(\tilde{\lambda}) = -j.
\label{eq:Itransform}
\end{align}
To connect Eqs.~\eqref{eq:phitransform} and~\eqref{eq:Itransform}, we note that the derivatives of $g$ are related to those of $\psi$ since $g$ and $\psi$ are inverses, $g_\pm(\psi_\pm(\tilde{\lambda})) = \tilde{\lambda}$.
Differentiating both sides of this equation and rearranging gives $g_\pm'(\psi_\pm(\tilde{\lambda})) = 1 / \psi_\pm'(\tilde{\lambda})$.
Note that the condition defining $\tilde{\mu}$ in Eq.~\eqref{eq:phitransform}, $g_\pm'(\tilde{\mu}) = \mp t$, can now be expressed as a condition on $\psi$: when $\tilde{\mu} = \psi_\pm(\tilde{\lambda})$, then $\psi_\pm'(\tilde{\lambda}) = \mp 1/t$.
Inserting this back into Eq.~\eqref{eq:phitransform} gives
\begin{align}
\nonumber \phi_\pm(t) &= \mp g_\pm(\psi_\pm(\tilde{\lambda})) - \psi_\pm(\tilde{\lambda}) t, \text{     where }\psi_\pm'(\tilde{\lambda}) = \mp t^{-1}\\
\nonumber &= - \psi_\pm(\tilde{\lambda}) t \mp \tilde{\lambda}, \text{     where }\psi_\pm'(\tilde{\lambda}) = \mp t^{-1}\\
\nonumber &= t \left(- \psi_\pm(\tilde{\lambda}) \mp \tilde{\lambda} t^{-1} \right), \text{     where }\psi_\pm'(\tilde{\lambda}) = \mp t^{-1}\\
&= t I(\pm 1 / t),
\end{align}
with the last line following from Eq.~\eqref{eq:Itransform}.

\section{Two-state, three-pathway model}

Analytical forms for $\psi, I, g_{\pm},$ and $\phi_{\pm}$ can be found for the two-state, three-pathway model of the main text.
We take $d_{12}^{\rm rxn} = 1$, $d_{21}^{\rm rxn} = -1$ and $d_{12}^{\rm therm} = d_{21}^{\rm therm} = d_{12}^{\rm fuel} = d_{21}^{\rm fuel} = 0$.
Thus we monitor the rate of net current from reactant to products, which has a steady-state value
\begin{equation}
\left<j\right> = (\beta - \alpha) / S,
\end{equation}
where
\begin{align}
S &= k^{\rm therm}_{12} + k^{\rm therm}_{21} + k^{\rm fuel}_{12} + k^{\rm fuel}_{21} + k^{\rm rxn}_{12} + k^{\rm rxn}_{21},\\
\alpha &= k^{\rm rxn}_{21} (k^{\rm therm}_{12} + k^{\rm fuel}_{12}),\\
\beta &= k^{\rm rxn}_{12} (k^{\rm therm}_{21} + k^{\rm fuel}_{21}).
\end{align}

The tilted rate matrix for this reactant to product current is
\begin{equation}
\mathbb{W}(\lambda) = \begin{pmatrix} - k_{21}^{\rm rxn} - k_{21}^{\rm therm} - k_{21}^{\rm fuel} & k_{12}^{\rm rxn} e^{-\lambda} + k_{12}^{\rm therm} + k_{12}^{\rm fuel}\\
k_{21}^{\rm rxn} e^{\lambda} + k_{21}^{\rm therm} + k_{21}^{\rm fuel} & -k_{12}^{\rm rxn} - k_{12}^{\rm therm} - k_{12}^{\rm fuel}
\end{pmatrix}.
\end{equation}
The scaled cumulant generating function (SCGF) for current is found as the maximum eigenvalue of $\mathbb{W}(\lambda)$:
\begin{equation}
\psi(\lambda) = -\frac{S}{2} + \frac{1}{2} \sqrt{S^2 + 4\left(1 - e^{-\lambda}\right)\left(\alpha e^{-\lambda} - \beta\right)}.
\end{equation}
In this case, $\psi^{-1} = g$ can be computed analytically.
As clear from Fig. 2 of the main text, the inversion requires us to define a ``$+$'' and ``$-$'' branch of $g$:
\begin{equation}
g_\pm(\mu) = \ln \left(\frac{\alpha + \beta + S \mu + \mu^2 \pm \sqrt{(\alpha + \beta + S \mu + \mu^2)^2 - 4 \alpha \beta}}{2 \alpha}\right).
\end{equation}
Using the G{\"a}rtner-Ellis theorem, we compute $I$ and $\phi_{\pm}$ with Legendre-Fenchel transforms,
\begin{align}
I(j) &= -\min_\lambda \left(\psi(\lambda) + \lambda j\right)\\
\phi_\pm(t) &= \mp \min_\mu \left(g_\pm(\mu) \pm \mu t\right).
\end{align}
For this two-state model, the minimizations can be carried out analytically with a moderate amount of algebra.
For compactness, we define two new functions:
\begin{equation}
\gamma(j) = 2 + \sqrt{4 + j^{-2}\left(S^2 - 4(\alpha + \beta) + 4 \alpha \beta j^{-2}\right)}
\end{equation}
and
\begin{equation}
\delta(j) = \sqrt{\left(S^2 - 4(\alpha + \beta)\right)j^{-2} + 4 \gamma(j)}.
\end{equation}
In terms of $\gamma$ and $\delta$ we find the rate functions:
\begin{widetext}
\begin{equation}
I(j) = \begin{cases}
\frac{j}{2} \left(\frac{S}{j} - \delta(j) + 2 \ln (2 \alpha j^{-2}) - 2 \ln [\gamma(j) - \delta(j)]\right),& j \geq 0\\
-\frac{j}{2} \left(-\frac{S}{j} - \delta(j) + 2 \ln (2 \alpha j^{-2}) - 2 \ln [\gamma(j) - \delta(j)]\right) + \ln (\alpha / \beta),& j < 0
\end{cases},
\end{equation}
\begin{equation}
\phi_+(t) = \frac{1}{2} \left(S t - \delta(t^{-1}) + 2 \ln(2 \alpha t^2) - 2 \ln[\gamma(t^{-1}) - \delta(t^{-1})]\right).
\end{equation}
and
\begin{equation}
\phi_-(t) = \phi_+(t) + \ln(\beta / \alpha)
\end{equation}
\end{widetext}
Observe that this final equation agrees with Eq.\ (11) of the main text, where $\lambda^* = \ln(\alpha / \beta)$.
As discussed in the main text, the fact that $\psi_+(t)$ and $\psi_-(t)$ have identical $t-$dependence is a consequence of the symmetry $\psi_+(\lambda) = \psi_-(\lambda^* - \lambda)$.

\end{document}